\documentclass[journal,hyphens]{IEEEtran}

\newif\ifReviewerview 
\Reviewerviewfalse

\usepackage{graphicx}
\usepackage{dcolumn}
\usepackage{bm}
\usepackage{amsmath}
\usepackage[range-phrase = --, range-units = single]{siunitx} 
\usepackage{cite}

\usepackage[switch,columnwise]{lineno}

\usepackage[hidelinks]{hyperref}
\usepackage{orcidlink}

\usepackage{mathptmx}
\usepackage{etoolbox}
\usepackage[normalem]{ulem}

\ifReviewerview
    \newcommand{\rem}[1]{\textcolor{red}{V0: #1}} 
     
    \linenumbers
\else
        \newcommand{\rem}[1]{\textcolor{black}{}} 
         
\fi 

\usepackage[nameinlink,capitalize]{cleveref}
\makeatletter
\begingroup
\catcode`\_=\active
\protected\gdef_{\@ifnextchar|\subtextit\subtextup}
\endgroup
\def\subtextit|#1|{\sb{#1}}
\def\subtextup#1{\sb{\mathrm{#1}}}
\AtBeginDocument{\catcode`\_=12 \mathcode`\_=32768}
\makeatother
\begin{document}

\title{\numproduct{8x8} Patch-Antenna-Coupled TeraFET Detector Array for Terahertz Quantum-Cascade-Laser Applications}
\author{Jakob~Holstein~\orcidlink{0009-0001-8744-475X},
Nicholas~K.~North~\orcidlink{0009-0002-8221-2893},
Michael~D.~Horbury~\orcidlink{0000-0001-8235-8142},
Sanchit~Kondawar~\orcidlink{0000-0003-3918-9472}, 
Imon~Kundu~\orcidlink{0000-0002-3564-1903},  
Mohammed~Salih~\orcidlink{0009-0001-6882-4642},
Anastasiya~Krysl~\orcidlink{0009-0000-9227-2756},
Lianhe~Li~\orcidlink{0000-0003-4998-7259},
Edmund~H.~Linfield~\orcidlink{0000-0001-6912-0535},
Joshua~R.~Freeman~\orcidlink{0000-0002-5493-6352},
Alexander~Valavanis~\orcidlink{0000-0001-5565-0463}, 
Alvydas~Lisauskas~\orcidlink{0000-0002-1610-4221
}, \IEEEmembership{Member, IEEE},
and
Hartmut~G.~Roskos~\orcidlink{0000-0003-3980-0964}
\thanks{
This work was supported by the German Research Foundation (DFG) project RO 770 49-1 of the INTEREST priority program (SPP~2314); UK Research and Innovation (Future Leader Fellowship MR/S016929/1); and the Engineering and Physical Sciences Research Council (EPSRC), U.K. (Programme grant EP/W012472/1).
A.Lisauskas acknowledges funding received from the Lithuanian Science Foundation (project No. S-MIP-22-83).
For the purpose of open access, the author has applied a Creative Commons Attribution (CC BY) license to any Author Accepted Manuscript version arising from this submission.
This manuscript builds upon preliminary work presented at conferences, DOI: 
10.1109/IRMMW-THz57677.2023.10298916,
10.1109/IRMMW-THz57677.2023.10298938 and
10.1109/IRMMW-THz57677.2023.10299121.}

\thanks{J. Holstein, A. Krysl and H. G. Roskos are with the Physikalisches Institut, Johann Wolfgang Goethe-Universität, DE-60438 Frankfurt am Main, Germany (e-mail: holstein@physik.uni-frankfurt.de; roskos@physik.uni-frankfurt.de)}
\thanks{A. Lisauskas is with (i) the Institute of Applied Electrodynamics and Telecommunications, Vilnius University, LT-10257 Vilnius, Lithuania, and (ii) Physikalisches Institut, Johann Wolfgang Goethe-Universität, DE-60438 Frankfurt am Main, Germany}
\thanks{N.~K.~North,
M.~D.~Horbury,
S.~Kondawar,
I.~Kundu,
M.~Salih,
L.~Li,
J.~R.~Freeman,
A.~Valavanis,
and E.~H.~Linfield are with the School of Electronic and Electrical Engineering, University of Leeds, Leeds LS7~3LF, United Kingdom (e-mail: N.K.North@leeds.ac.uk; a.valavanis@leeds.ac.uk)}
}

\date{\today}
\maketitle
\begin{figure}[tb]
  \centering
  \includegraphics[keepaspectratio, width=\columnwidth]{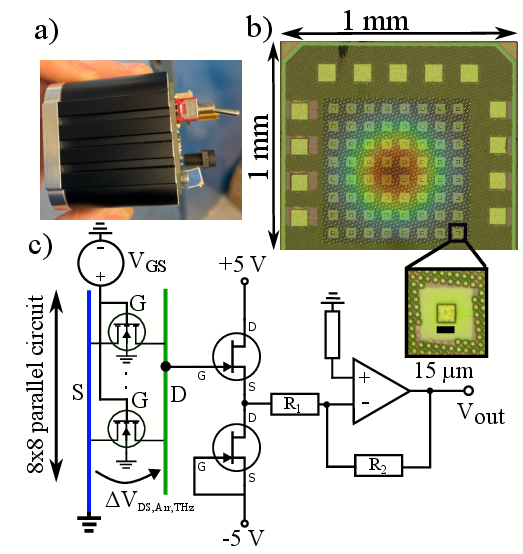}
  \caption{a)~Photograph of the detector package. b)~Microscope image of the multi-element detector consisting of $8\times 8$ identical patch-antenna coupled TeraFETs connected in a parallel readout-circuitry.  Inset 1: expanded view of individual antenna element. Inset 2: Visualization of the $8\times 8$ array configuration transposed onto an analytically calculated Gaussian beam focus providing $1/e^2$-diameter of \SI{520}{\micro \meter}. 
  c)~Schematic diagram of the $8\times 8$ rectifying $n$-channel MOSFET elements in a parallel configuration. The rectified THz-signal $V_{DS,Arr,THz}$ is fed into an amplifier circuit providing a gain of approx. \num{70}.}
  \label{fig:Overview_Detector_IVdata_compSP}
\end{figure}

\begin{abstract}
Monolithically integrated, antenna-coupled field-effect transistors (TeraFETs) are rapid and sensitive detectors for the terahertz range (0.3-10~THz) that can operate at room temperature. We conducted experimental characterizations of a single patch-antenna coupled TeraFET optimized for 3.4~THz operation and its integration into an 8×8 multi-element detector configuration. In this configuration, the entire TeraFET array operates as a unified detector element, combining the output signals of all detector elements. Both detectors were realized using a mature commercial Si-CMOS 65-nm process node. Our experimental characterization employed single-mode Quantum-Cascade Lasers (QCLs) emitting at 2.85~THz and 3.4~THz. The 8x8 multi-element detector yields two major improvements for sensitive power detection experiments. First, the larger detector area simplifies alignment and enhances signal stability.
Second, the reduced detector impedance enabled the implementation of a TeraFET+QCL system capable of providing a -3~dB modulation bandwidth up to 21~MHz, which is currently limited primarily by the chosen readout circuitry.
Finally, we validate the system's performance by providing high resolution gas spectroscopy data for methanol vapor around 3.4~THz, where a detection limit of 1.6e-5 absorbance, or 2.6e11~molecules/cm$^3$ was estimated under optimal coupling conditions.

\end{abstract}

\IEEEpeerreviewmaketitle

\section{Introduction}
\label{sec:Introduction}

\IEEEPARstart{C}{ontinuously} growing research activities are evident in the terahertz frequency range of the electromagnetic spectrum (\SIrange{0.3}{10}{\tera \hertz}). 
Promising technologies and applications span from high-bandwidth communications to spectroscopy of gases and solids, which exhibit characteristic spectroscopic absorption lines in the THz band~\cite{hubers_high-resolution_2019}.
For narrow-band THz applications, a combination of powerful coherent sources, and fast low-noise detectors is typically required.
Significant progress has been made in recent years in source technology with the introduction of terahertz quantum-cascade-lasers (QCL).
These serve as powerful terahertz sources, achieving continuous wave (cw) power levels on the order of milliwatts  or peak power up to the watt-level in pulsed operation~\cite{li_terahertz_2014}, through intersubband transitions in a semiconductor heterostructure~\cite{williams_terahertz_2007}.
QCLs offer high spectral purity, and are tunable over a range of several GHz and their design frequencies can be selected across the entire $\sim$2--5\,THz range.
QCLs have been applied in a wide range of applications include gas-phase molecular spectroscopy~\cite{richter_direct_2021,wubs_terahertz_2023,cuisset_terahertz_2021,hubers_high-resolution_2019}, imaging~\cite{dean_terahertz_2014}, metrology, and numerous other applications \cite{vitiello_terahertz_2022,gao_recent_2023}.
In addition to powerful terahertz sources, advancements in terahertz detectors, such as those in the field of antenna-coupled field-effect transistors (TeraFETs), have demonstrated significant progress in recent years.
TeraFETs are fast and sensitive terahertz detectors, achieving optical, non-area normalized noise equivalent power (NEP) values on the order of \SI{20}{\pico\watt\per\sqrt\hertz} in the frequency range below 1~THz~\cite{javadi_sensitivity_2021}.
They have also been shown to detect radiation up to 30~THz~\cite{regensburger_picosecond-scale_2019}.
In contrast to competing room-temperature terahertz detector technologies, such as Schottky diodes~\cite{yadav_state---art_2023}, TeraFETs provide NEP well below \SI{500}{\pico \watt \per \sqrt \hertz} above 3~THz~\cite{zdanevicius_field-effect_2018}.
They have been utilized as sensitive power detectors in conjunction with QCLs for power detection at 4.75~THz~\cite{zdanevicius_field-effect_2018}, or as efficient detectors for pulsed QCL emission at 3~THz \cite{ikamas_efficient_2017}. Moreover, they have been employed in terahertz scanning near-field optical microscopy (s-SNOM) applications~\cite{wiecha_antenna-coupled_2021}, and for passive detection of thermal radiation~\cite{cibiraite-lukenskiene_passive_2020}.
TeraFETs are realized in a variety of material technologies~\cite{javadi_sensitivity_2021} such as graphene~\cite{zak_antenna-integrated_2014, generalov_400-ghz_2017,ludwig_terahertz_2024}, AlGaN/GaN~\cite{bauer_high-sensitivity_2019}, AlGaAs/GaAs~\cite{regensburger_picosecond-scale_2019}  or Si-CMOS~\cite{boppel_cmos_2012, bauer_antenna-coupled_2014,ludwig_modeling_2024}. 
In addition to sensitive room-temperature operation, they are frequency selective within the terahertz range.
This selectivity, directed towards specific regions of the terahertz range, can be effectively modulated through the utilization of either broadband antennas~\cite{ikamas_broadband_2018} or narrow-band antennas such as patch-antennas~\cite{bauer_antenna-coupled_2014, ikamas_all-electronic_2021}.
Additionally, their fabrication process is compatible with standard foundry processes, leading to lower production costs and increased reproducibility due to the available process maturity.
Among the available technologies, process maturity of Si-CMOS is most advanced.
In this work, we present a study of a TeraFET design leveraging the advanced process maturity of the Taiwan Semiconductor Manufacturing Company (TSMC) 65\mbox{-}nm process node.
We present our advancements in developing a highly sensitive single patch antenna-coupled TeraFET tailored for \SI{3.4}{\tera \hertz} spectroscopy applications, alongside its incorporation into an array geometry, comprising $8\times 8$ TeraFET elements.
Both the individual component and the array configuration underwent experimental characterization employing terahertz Quantum Cascade Lasers (QCLs).
Our multi-element detector approach is strategically optimized to achieve a sufficiently expansive active detector area, coupled with high readout bandwidth, while maintaining a low noise level.
This design addresses a critical challenge within the realm of gas spectroscopy applications, where conventional thermal detectors, such as bolometers, often exhibit slow response times.
Currently, this limitation typically restricts applications, such as gas sensing or communications, to modulation rates not exceeding \SI{1}{\kilo \hertz}, as recently addressed in Ref.~\cite{wubs_terahertz_2023}. In contrast, the rapid intrinsic electronic detection mechanism in TeraFETs, involving plasmon excitation/distributed resistive mixing within a 2D-electron-gas (2DEG) coupled to asymmetric antenna elements~\cite{dyakonov_shallow_1993, dyakonov_detection_1996}, enables modulation bandwidths spanning from the low Hz-regime to the GHz-range~\cite{regensburger_picosecond-scale_2019}.
In \autoref{fig:Overview_Detector_IVdata_compSP}, the detector system implemented in this study is shown. It offers a -3~dB modulation bandwidth of 15~MHz around the most sensitive bias point, with the potential to extend up to 21~MHz at the expense of sensitivity (c.f. ~\autoref{fig:BodePlot}).
 Presently, the modulation bandwidth is primarily constrained by the chosen amplifier circuit.
Nevertheless, it has facilitated time-resolved gas spectroscopy experiments, as demonstrated in \cite{horbury_real-time_2023}. Moreover, its applicability for high-bandwidth QCL-characterization experiments has been validated through the direct observation of the RF-modulation induced threshold-shift~\cite{north_real-time_2023}.

\begin{figure}[h]
  \centering
    \includegraphics[width=\columnwidth]{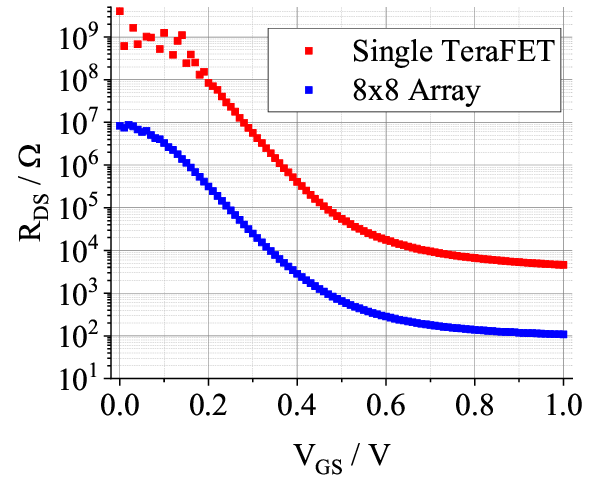} \\
    \caption{Experimental $R_{DS}(V_{GS})$-characteristic determined for single TeraFET and 8x8 array detector.}
    \label{fig:RDS_analysis}
\end{figure}

In a previous study~\cite{zdanevicius_field-effect_2018}, patch-antennas for spectroscopy applications from \SIrange{3}{4.7}{\tera \hertz} reached a minimum area normalized NEP of \SI{404}{\pico\watt\per\sqrt\hertz} at \SI{4.75}{\tera \hertz}.
Here, we present experimental studies conducted on both single element detector and a detector array, utilizing parallel readout circuitry to combine signals from all rectifying detector elements.
Our experimental results were characterized using \SI{2.85}{\tera \hertz} and \SI{3.4}{\tera \hertz} single mode QCLs to provide measurements both close to and far from the expected antenna resonance.
\section{Detector Implementation}
Two detectors, namely the single-patch antenna coupled TeraFET and the 8x8 detector array, were developed and investigated for this study. The single-patch antenna coupled TeraFET serves as the unit cell of the 8x8 array detector. Therefore, meticulous attention was paid to the design of the single element.
In the unit cell element, the rectifying transistor was linked to a square patch antenna measuring \qtyproduct[product-units=power]{15x15}{\micro\meter}. 
For this purpose, the patch-antenna was simulated within the environment of the commercial 65-nm stack applied for fabrication  to determine frequency-dependent antenna impedances. During the simulations, the impedance was optimized for best performance at 3.5~THz with an expected typical spectral width FWHM $\approx.~8-10\% \cdot f_{res}$. 
\footnote{A spectral bandwidth of FWHM $\approx$ 8-10~\% of the resonant frequency $f_{\text{res}}$ is consistent with our results on previous designs with resonant frequencies up to 2.5~THz. However, preliminary data obtained by Dmytro B. But at CENTERA Laboratories, Warsaw, Poland, examining the antenna characteristics of the single-patch detector coupled to a superstrate Si-lens \cite{krysl_control_2022}, using thermal interferometry, indicates discrepancies for our detector. In this dataset, we observe a FWHM of approximately 1~THz, as well as a maximum signal at 3.0~THz. 
This preliminary finding could support the results presented in the results chapter. However, direct comparisons are not feasible as the resonant frequency shifts to lower frequencies due to the use of the superstrate lens. In the case of the presented detector, we expect this effect to be less pronounced compared to \cite{krysl_control_2022} as the patch is placed in a lower metal layer M8 (groundplane-to-patch distance \SI{5.4}{\micro \meter}).}
Following the simulation of antenna impedances, the transistor dimensions (gate length \SI{60}{\nano \meter}, transistor channel width \SI{200}{\nano \meter}) were chosen. This selection process, conducted using Keysight PathWave Advanced Design System (ADS) software
\footnote{Link to Keysight PathWave Advanced Design System (ADS) 2023: \url[Link]{https://www.keysight.com/de/de/products/software/pathwave-design-software/pathwave-advanced-design-system.html}}, aimed to achieve lowest NEP performance. \footnote{A detailed overview on our design and modeling procedure is given in \cite{ludwig_modeling_2024}.} 

\autoref{fig:RDS_analysis} illustrates the electrical IV characteristics of both the single TeraFET element and the 8x8 array detector.

The individual transistors were linked to a square patch antenna measuring \qtyproduct[product-units=power]{15x15}{\micro\meter}. This dimension was found to provide best performance at the design frequency, while embedded in the commercial 65-nm foundry fabrication.

The first detector configuration comprised a single patch-antenna coupled transistor, while the second configuration featured an array of $8\times 8$ transistors connected in parallel, with a pitch of \SI{75}{\micro\meter} between neighboring antenna centers, as shown in \autoref{fig:Overview_Detector_IVdata_compSP}.
To account for the anticipated low detector noise, the drain-output signal from each transistor is routed into a low-noise buffer stage (Source-Follower with JFET Current Source).
This stage comprises an N-Channel JFET transistor pair selected for their combination of low voltage-noise and low input capacitance (\SI{25}{\pico\farad}). The specific transistor model used for this purpose is the Linear Systems LSK 389B \cite{linear_systems_lsk389_nodate}.
This configuration is expected to provide unity gain. However, an amplitude ratio of approx. $\num{0.7}$ was determined.
The buffer stage's output voltage is directed into an inverting voltage amplifier circuit featuring a high gain-bandwidth product of \SI{1.5}{\giga\hertz}. The specific model used for this amplifier is the Texas Instruments LMH-6624.
We chose $R_2 / R_1 = 100$, resulting in a system voltage amplification ratio (Gain) $V_{out} / V_{in} = -70$. 
\footnote{The single patch detector's amplifier circuit did not use the buffer stage. Therefore, a voltage amplification ratio of \num{-100} was taken into account.}

The gate-bias-dependent source-drain resistance, $R_{DS}(V_{GS})$ was  measured experimentally for both detectors by sweeping the gate-source voltage while maintaining a fixed $ V_{DS}=\pm \SI{10}{\milli \volt}$. The recorded values are depicted in \autoref{fig:RDS_analysis}.
As anticipated from fundamental principles such as Kirchhoff's law, the resistance of the multi-element detector decreases from approximately \( R_{DS,SP} \approx \SI{16}{\kilo\ohm} \) (single element) to \( R_{DS,Arr} \approx \SI{300}{\ohm} \) at the expected bias operational point around \( V_{GS} = \SI{0.6}{\volt} \) (see Section~\ref{sec:experimental_results}).
This reduction in output impedance simplifies the connection of the multi-element detector to an external amplifier network, rendering it more compatible than a single-element detector and obviating the need for specialized impedance matching techniques.
In addition to the reduced impedance previously discussed, the presented detector features an increased sensitive area, significantly simplifying the alignment procedure in the experimental beam focus.
\section{Theoretical background}

\subsection{Figure of merit - Noise-Equivalent-Power (NEP)}

The most pertinent figure of merit for detectors is the noise equivalent power (NEP), which characterizes the minimum optical power required to produce a detector signal surpassing the noise level within a specified measurement bandwidth, $\Delta  f$. 
For TeraFETs, under an unbiased drain-source condition\footnote{Unbiased drain-source condition: No external DC-biasing over the transistor channel during rectification process.}, the primary noise contribution arises from thermal Johnson-Nyquist noise \cite{bauer_antenna-coupled_2014}.
The noise voltage spectral density can thus be determined using
\begin{equation}
    \langle v_{\mathrm{N}} \rangle = \sqrt{4k_{\mathrm{B}} T R_{\mathrm{DS}} (V_{\mathrm{GS}})}
    \label{eq:JN_noise}
\end{equation}
where $k_{\mathrm{B}}$ is the Boltzmann constant and $T=\SI{293}{\kelvin}$ is the temperature of the detector (here operated at room-temperature).
Following \autoref{eq:JN_noise}, the detector's noise contribution is directly proportional to the square-root of the drain--source resistance of the channel, $R_{\mathrm{DS}}(V_{\mathrm{GS}})$, which is controlled by the applied gate--source voltage (\emph{cf.}
~\autoref{fig:RDS_analysis}).

\begin{figure}[h]
  \centering
   \includegraphics[width=\columnwidth]{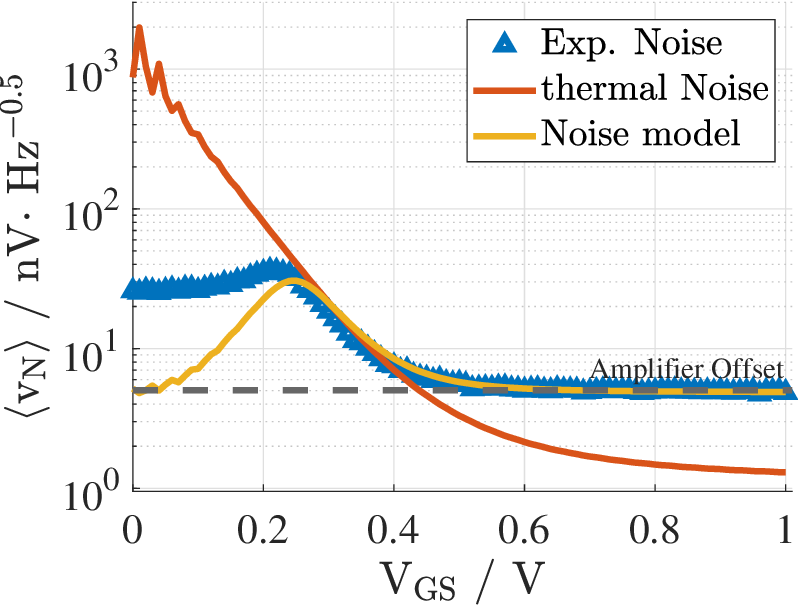} \\
    \caption{Experimental Noise-spectral density for 8x8 array detector integrated in an amplifier circuit determined at $f_{mod}$=\SI{50}{\kilo\hertz} using AMETEK 7265 lock-in amplifier. Here, it is shown corrected for the amplifier circuits voltage-gain to allow comparison with thermal Johnson-Nyquist noise contribution calculated from \autoref{eq:JN_noise} (red). In addition, yellow curve models total system noise contributions by consideration of a first order low pass-filter and an offset resulting from the applied amplifier circuitry.\protect\footnotemark}
    \label{fig:noise_analysis}
\end{figure}

However, since the detector is included in an amplifier circuit, statistically independent noise contributions $\langle v_k \rangle$ must be taken into consideration~\cite{brown_fundamentals_2003}
\begin{equation}
    \langle v_{\mathrm{ges}} \rangle^2 = 4k_{\mathrm{B}} T R_{\mathrm{DS}} (V_{\mathrm{GS}})+ \sum_{k=1}^N\langle v_k \rangle^2
    \label{eq:JN_noiseplus}
\end{equation}
While \autoref{eq:JN_noise} is valid for the single TeraFET element, where the experimental noise spectral density agrees well with the predicted value, the 8x8 detector network's $V_{GS}$-dependent noise-spectral density cannot fully be described by \autoref{eq:JN_noise}. Due to the strongly reduced $R_{DS}$ of the 8x8 network (cf. \autoref{fig:RDS_analysis}). 
In \autoref{fig:noise_analysis}, the detector system's (8x8 TeraFET + readout circuit) experimental noise spectral density is shown as a function of $V_{GS}$. As visible in \autoref{fig:noise_analysis} the Johnson Nyquist  prediction is underestimating experimental noise spectral density  for $V_{GS}>\SI{0.5}{\volt}$ as noise contributions such as the applied buffer or amplifier elements add $V_{GS}$-independent noise becoming dominant contribution in this regime, which can be  explained via \autoref{eq:JN_noiseplus}.

\footnotetext{The simple model considered here assumes a noise spectral density transfer-function via $\langle \mathrm{v}_{\mathrm{Ges}} \rangle^2 = \langle \mathrm{v}_{\mathrm{Amp}} \rangle^2 + \left(\frac{\langle \mathrm{v_N} \rangle}{\sqrt{1+\mathrm{(\omega R_{DS} C })}^2}\right)^2$. Here $\langle \mathrm{v}_{\mathrm{Amp}} \rangle \approx \SI{4.7}{\nano \volt \per \sqrt{\hertz}}$ summarizes amplifier plus buffer plus background noise. Second term summarizes network Johnson-Nyquist contribution passing  a $V_{GS}$-dependent $RC$-lowpass, where we use the  detector resistance $R_{DS}(V_{GS})$ and  $C \approx \SI{25}{\pico \farad}$, which is the input capacitance of the JFET buffer cp. \autoref{fig:Overview_Detector_IVdata_compSP}.}

Noise-Equivalent-Power (NEP) is then obtained via
\begin{equation}
    \text{NEP} = \frac{\langle v_{\mathrm{ges}}\rangle}{\mathcal{R}_{\mathrm{V}}} = \frac{P_{\mathrm{THz}}}{\text{SNR}\sqrt{\Delta f}}
    \label{eq:NEP}
\end{equation}
where $\mathcal{R}_{\mathrm{V}}$ denotes the detector's voltage responsivity.
It directly relates incident terahertz power $P_{THz}$ to the signal-to-noise-ratio (SNR), weighted by the applicable measurement bandwidth, here determined by the applied lock-in amplifier integration time.
Given that effective antenna area $A_{\mathrm{eff}}$ in case of a single patch-antenna is small compared to the focused beam, a valid measure to compare performance between differing antennas is the cross-sectional responsivity ($\mathcal{R}_{\mathrm{V},\mathrm{CS}}$) (\emph{cf}.~\cite{ikamas_all-electronic_2021}).
This can be determined experimentally by integrating the THz-rectified drain--source voltage as the detector is scanned across the beam focal plane:

\begin{equation}
    \mathcal{R}_{\mathrm{V},\mathrm{CS}} = \frac{1}{A_{\mathrm{eff}}P_{\mathrm{THz}}}  \iint \Delta V_{\mathrm{DS}}(x,y)\, \text{d}x\, \text{d}y
    \label{eq:RV_Cross-sectional}
\end{equation}
We approximate the effective antenna area, based on antenna simulations~\cite{zdanevicius_field-effect_2018} with 
 \begin{equation}
    A_{\text{eff}} = \frac{D \lambda_0^2}{4\pi}
    \label{eq:Effective_area_Directivity}
\end{equation}
where $D$= 5.6~dBi represents simulated antenna directivity and $\lambda_0(\SI{3.4}{\tera \hertz})=\SI{88}{\micro \meter}$ is the wavelength of the incident radiation in free-space.
The resulting effective antenna area is \qtyproduct[product-units=power]{2250}{\micro \meter \squared}.

\subsection{Performance parameters of detector network}
The overall THz rectified voltage $\Delta V_{\mathrm{DS},\mathrm{Arr}}$ in the parallel network is obtained from 64 identical, rectifying elements.
Due to the non-homogeneous intensity distribution in the experimental focus, the incoming THz power $P_{\text{THz},k}$ coupled to the array's $k^{\mathrm{th}}$ element varies considerably.
In the  general case of $N$ identical rectifying elements arranged in a parallel circuit, Thévenin’s theorem provides the overall rectified voltage signal as
\begin{eqnarray}
    \Delta V_{\mathrm{DS},\mathrm{Arr}} &=& \frac{\mathcal{R}_{\mathrm{V},SP}\left(\nu_{\mathrm{THz}}\right)}{N}\sum_{k=1}^{N}P_{\mathrm{THz},k}
    \label{eq:derivation_responsivityArr}
\end{eqnarray}
where $\mathcal{R}_{\mathrm{V},SP}$ is the cross-sectional responsivity of a single-patch-antenna device.
We can determine the optical responsivity of the detector array by employing the assumption that $P_{THz}$ is approximately equal to the summation of individual THz powers $P_{\mathrm{THz},k}$ for all elements in the array, i.e. $P_{THz} \approx \sum_{k=1}^{N}P_{\mathrm{THz},k}$. 
It therefore follows that the responsivity of an $N$-element array is reduced by $N$ compared to the single detector's cross-sectional responsivity:
\begin{equation}
    \mathcal{R}_{V,Arr,opt} = \frac{\mathcal{R}_{V,SP}}{N} 
    \label{eq:Responsivityrelation}
\end{equation}
where $\mathcal{R}_{V,Arr,opt}$ denotes the $N$-element array detector's optical responsivity.\footnote{In the literature, optical responsivity simply relates rectified signal to the incident terahertz power without further normalization~\cite{javadi_sensitivity_2021}.} 
Following \autoref{eq:JN_noise}, the optical NEP for the array device is given by:
\begin{equation}
    \text{NEP}_{Arr,opt} = \text{NEP}_{SP,CS}\cdot \sqrt{N}
    \label{eq:NEPArrCS}
\end{equation}
In summary, the implementation of the described array detector leads to reduced sensitivity, manifested in an increased $\text{NEP}_{\text{Arr,opt}}$. However, this configuration can be optimized to achieve high readout speed and a larger effective antenna area. In addition to performance parameters, the enlarged detector area simplifies detector handling and enhances signal stability concerning vibrations or beam jitter.

\section{Experimental Results}
\label{sec:experimental_results}

\begin{figure}[h]
  \centering
  \includegraphics[keepaspectratio, width=0.9\columnwidth]{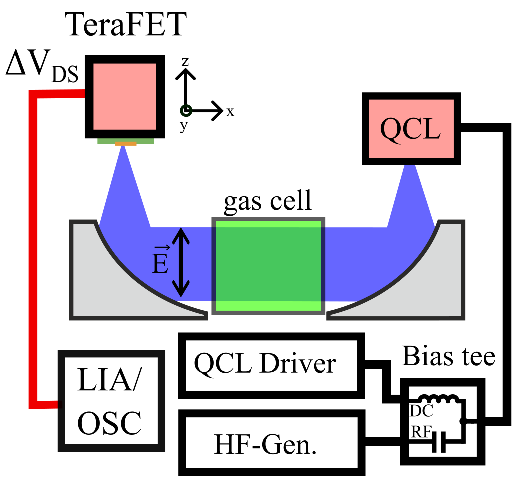}
   \caption{Sketch of experimental measurement setup applied for TeraFET detector characterizations and as well as gas spectroscopy experiments. For spectroscopy experiments (compare \autoref{fig:MethanolSpectrum}), a \SI{95}{\centi \meter}-gas cell was is placed in the collimated beam path.}
  \label{fig:Measurement_setup}
\end{figure} 

Terahertz response of each detector was characterized using the measurement apparatus illustrated in \autoref{fig:Measurement_setup}.
For each measurement, a QCL was mounted in a closed-cycle ColdEdge\texttrademark{} CH204N cryocooler at a temperature of \SI{20}{\kelvin}, and driven electronically in continuous-wave mode using a Wavelength Electronics\texttrademark{} QCL1000 LAB current source.
The THz output from the QCL was collected and collimated using an off-axis paraboloidal gold mirror (2", $f/1.5$) and focused onto the detector element using a second identical mirror.
The detectors under test were mounted on a motorized three-dimensional linear translation stage to enable accurate positioning in the experimental beam focus, and automated beam scanning.
A sinusoidal modulation ($f_{mod}=\SI{150}{\kilo \hertz}$, \SI{2}{\volt_{pp}}) was applied to the QCL via a bias-T, utilizing an RF sweeper. Subsequently, the detector signal was demodulated and recorded using a Signal Recovery\texttrademark{} 7265 lock-in amplifier.

\subsection{THz QCL beam analysis}

\begin{figure}[tb]
  \centering
  \includegraphics[keepaspectratio, width=\columnwidth]{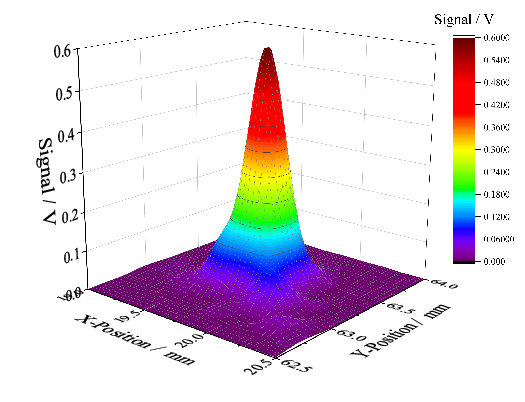}
  \caption{Experimental 3.4\,THz QCL beam intensity imaged across the focal plane using single patch coupled TeraFET mounted on a motorized three-dimensional linear translation stage. The increment in both $x$ and $y$ direction is \SI{20}{\micro \meter}.}
  \label{fig:Experimental_focus_SP}
\end{figure}
The THz beam profile shown in \autoref{fig:Experimental_focus_SP} was obtained using single patch-antenna coupled TeraFET ($V_{GS}=\SI{0.6}{\volt}$). 
Experimentally, the detector was scanned with a step-size of \SI{20}{\micro \meter} across the focal plane of a 3.4\mbox{-}THz QCL adapted from hybrid bound-to-continuum/resonant-phonon design~\cite{wienold_high-temperature_2014}, which was chosen to approximately match the resonance frequency of the patch antenna.
As visualized in \autoref{fig:Experimental_focus_SP}, our results follow a Gaussian intensity distribution showing an experimental $1/e^2$ diameter of \SI{520}{\micro \meter}.  
The area covered within the 1/$e^2$ beam diameter is approximately 54~times greater than the effective single-patch antenna area calculated using \autoref{eq:Effective_area_Directivity}, and hence only a small fraction of the beam power is incident upon the detector at a given time.
Although optical coupling to the single-patch antenna could be improved using a superstrate Si lens as introduced in Ref.~\cite{krysl_control_2022} the increased directivity would introduce very high sensitivity to system alignment.\footnote{The additional permittive load introduced by a superstrate lens also leads to a shift in the resonant towards smaller frequencies~\cite{krysl_control_2022}.}
As such, mechanical vibrations, or small drifts in optical alignment result in large changes in detector signal. 
By contrast, \autoref{fig:Overview_Detector_IVdata_compSP}(b) provides a visualization of the focused THz beam spot, transposed onto the  $8\times 8$ multi-element detector, showing that the full beam spot is captured directly by the array without the requirement for additional optical elements.

Based on the beam profile data shown in \autoref{fig:Experimental_focus_SP}, and the total focused power, $P_{THz}(\SI{3.4}{\tera \hertz})=\SI{3.6}{\milli \watt}$, the achievable cross-sectional responsivity of the single-patch detector was estimated according to \autoref{eq:RV_Cross-sectional} as $\mathcal{R}_{V,SP,CS} = \SI{285}{\volt \per \watt}$. 
This corresponds to a cross-sectional Noise-Equivalent-Power $\text{NEP}_{SP,CS} = \SI{57}{\pico\watt\per\sqrt\Hz}$, which is a state-of-the-art sensitivity for room-temperature electronic detectors at \SI{3.4}{\tera \hertz} (\emph{cp.}~\cite{zdanevicius_field-effect_2018}).
For the 8$\times$8 array configuration, we can estimate performance parameters according to \autoref{eq:derivation_responsivityArr} and \autoref{eq:NEPArrCS} with $\mathcal{R}_{V,Arr,opt} = \SI{4}{\volt \per \watt}$ and $\text{NEP}_{Arr,opt} \approx \SI{500}{\pico\watt\per\sqrt\Hz}$.
This illustrates that although the array geometry leads to a reduced effective responsivity, the optical NEP is expected to remain well below \SI{1}{\nano\watt\per\sqrt\hertz}. 

\subsection{Responsivity and NEP at 2.85~THz and 3.4~THz}

The performance of the array was measured directly, both close to resonance (3.4~THz, available power \SI{1.5}{\milli \watt}) and away from resonance, using a second QCL with 2.85~THz emission frequency, and $P_{THz}=\SI{0.75}{\milli \watt}$, which was adapted from the active-region design in Ref.~\cite{wienold_low-voltage_2009}.

\begin{figure}[tb]
  \centering
  \includegraphics[keepaspectratio, width=\columnwidth]{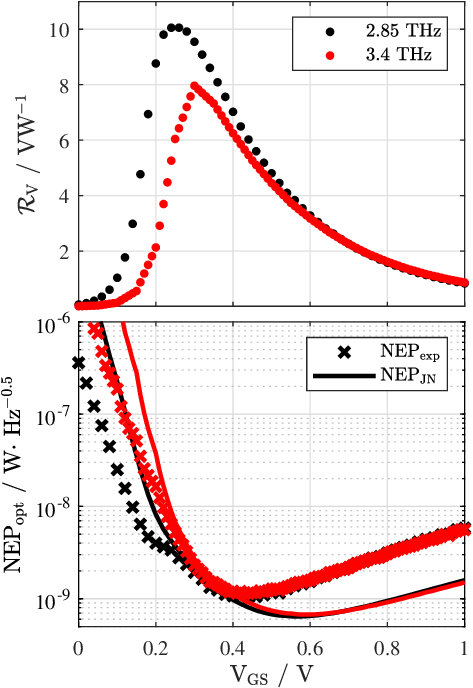}
  \caption{
  Performance characteristics of the detector array as a function of the applied $V_{GS}$ at \SI{2.85}{\tera \hertz} (black) and \SI{3.4}{\tera \hertz} (red). Top: Optical (non-area normalized) voltage responsivity $\mathcal{R}_{\mathrm{V}}$.\\ Bottom: Optical Noise-Equivalent-Power (NEP$_{opt}$). Here, we distinguish between $\mathrm{NEP}_{\mathrm{exp}}$ (Calculated based on measured system noise \autoref{fig:noise_analysis}) and $\mathrm{NEP}_{\mathrm{JN}}$ calculated via thermal Johnson-Nyquist prediction according to \autoref{eq:JN_noise}.}
  \label{fig:array_results2934thz}
\end{figure}

The values of responsivity and NEP were determined as a function of gate bias, as shown in \autoref{fig:array_results2934thz}.
The responsivity, at $V_{GS} = \SI{0.6}{V}$, was found to be 3.2~V/W for 3.4~THz and 3.3~V/W for 2.85~THz.
The former result is close to the estimate in the previous section, but
the higher responsivity at 2.85~THz is contrary to expectations regarding the resonance frequency.

The experimentally determined noise also deviates from a pure thermal Johnson-Nyquist noise prediction.
If the NEP is calculated solely on the basis of thermal Johnson-Nyquist noise (cp. \autoref{eq:JN_noise}), which conforms to the literature, the minimum optical NEP is found at $V_{GS} \approx \SI{0.57}{\volt}$.
At 2.85~THz it shows \SI{650}{\pico\watt\per\sqrt\Hz} and 
\SI{680}{\pico\watt\per\sqrt\Hz} for the 3.4~THz measurement.
However, the minimum experimental optical NEP is found at \SI{0.45}{V} and is \SI{1.03}{\nano\watt\per\sqrt\Hz} at 2.85~THz and \SI{1.15}{\nano\watt\per\sqrt\Hz} at 3.4~THz.
Due to the amplifier-noise offset indicated in \autoref{fig:noise_analysis}, the NEP at the previously predicted operating point (approx. 0.6~V) is therefore approximately \SI{1.6}{\nano\watt\per\sqrt\Hz} for both frequencies. 



\subsection{Modulation bandwidth}
\begin{figure}[tb]
  \centering
  \includegraphics[width=\columnwidth]{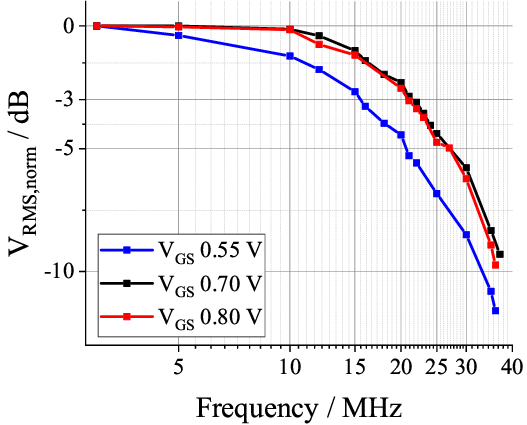}
  \caption{Experimentally determined normalized output voltage $V_{rms}$ response of the TeraFET array as a function of QCL modulation frequency. Results are shown for three $V_{GS}$ levels.}  
  \label{fig:BodePlot}
\end{figure}
Prior to testing the modulation bandwidth of the detector, it was necessary to determine the bandwidth limit of the measurement apparatus, which in this case was set by the QCL source.
The gain dynamics of QCLs are intrinsically very fast, enabling modulation at frequencies up to \SI{35}{\giga\hertz}~\cite{gellie_injection-locking_2010}.
However, in practice, this requires careful impedance matching between the drive electronics and the QCL ridge.
To characterize the QCL bandwidth, a modulation was applied using a Keysight\texttrademark{} RF sweeper, via a bias-T.
At low modulation frequencies, this has the effect of periodically raising the QCL above its lasing threshold.
As such, an apparent reduction in the threshold current of the laser is observed in time-averaged THz power measurements~\cite{north_real-time_2023}. 
Using this approach, the modulation bandwidth of the 2.85~THz QCL was determined to be $f_{-3dB,QCL} \approx \SI{100}{\mega \hertz}$, thus enabling accurate detector characterization within this limit.
The bandwidth of the TeraFET detector array was analyzed by recording the modulated QCL emission in the time-domain using a \SI{100}{\mega\hertz} Keysight DSOX2014A oscilloscope. 
\autoref{fig:BodePlot} shows the normalized $V_{rms}$-signal as a function of the modulation frequency for three different $V_{GS}$-potentials.
Experimentally, we found $f_{-3dB}(V_{GS}=\SI{0,55}{\volt})=\SI{15.5}{\mega \hertz}$, which is consistent with the expected $RC$-limitations. 
For $V_{GS}=\SI{0,7}{\volt}$ and \SI{0,8}{\volt}, we found $f_{-3dB} \approx \SI{21}{\mega \hertz}$. Here, the maximum modulation bandwidth is no longer primarily limited by the detector resistance as $R_{DS}$ decreases with increasing $V_{GS}$ (c.f. \autoref{fig:Overview_Detector_IVdata_compSP}), but is instead limited by the available amplifier bandwidth.
\footnote{Due to the chosen gain factor of  70, we expect approx. \SI{21,4}{\mega \hertz}. Transfer characteristics of the detector circuit simulated using TINA-TI \texttrademark{} software closely approximate the experimentally determined $f_{-3dB}$ bandwidths for all three cases.}

A close comparison of \autoref{fig:array_results2934thz} with \autoref{fig:BodePlot} reveals that the most sensitive operating point differs from the $V_{GS}$-bias providing fastest operating point.

\subsection{Gas spectroscopy application}
To illustrate the suitability of the detector for lab-based gas spectroscopy applications, we investigated the transmission spectrum of the 3.4~THz QCL through methanol (CH$_3$OH) vapor.
For these measurements, a \SI{0.95}{\meter} glass gas cell with 3-mm-thick poly-methyl-pentene Brewster-angle windows was placed in the terahertz beam path of the experimental setup (\emph{cf.} \autoref{fig:Measurement_setup}).
The gas cell was purged with nitrogen, evacuated using a scroll-pump, and methanol vapor (99.8\%, Sigma-Aldrich) was evaporated from a glass storage tube into the gas cell using an automated mass-flow controller (MKS, GE50A) to a pressure of \SI{0.5}{Torr} (\SI{66.6}{\pascal}).
The QCL emission frequency was chirped by applying a sawtooth modulation from 350~mA to 1000~mA to the laser bias at a 230\mbox{-}Hz repetition rate.
The THz power transmitted through the gas cell was sampled at a rate of \SI{150}{\kilo \hertz} using the $8\times 8$ detector array, and a National Instruments data acquisition board.
This provided measurements of the QCL power at 1~mA bias-steps across a   sweep time of \SI{4.5}{\milli \second}.
Signal-to-noise ratio of the resulting spectrum was improved by averaging over 13800 acquired spectra (i.e., a total measurement time of \SI{1}{minute}), resulting in an effective measurement bandwidth of \SI{10.9}{\hertz} and an equivalent optical noise-level of $\text{NEP}\cdot \sqrt{\Delta f}=\SI{5.3}{\nano\watt}$ for each spectral sampling point. 

\begin{figure}[tb]
  \centering
  \includegraphics[keepaspectratio, width=\columnwidth]{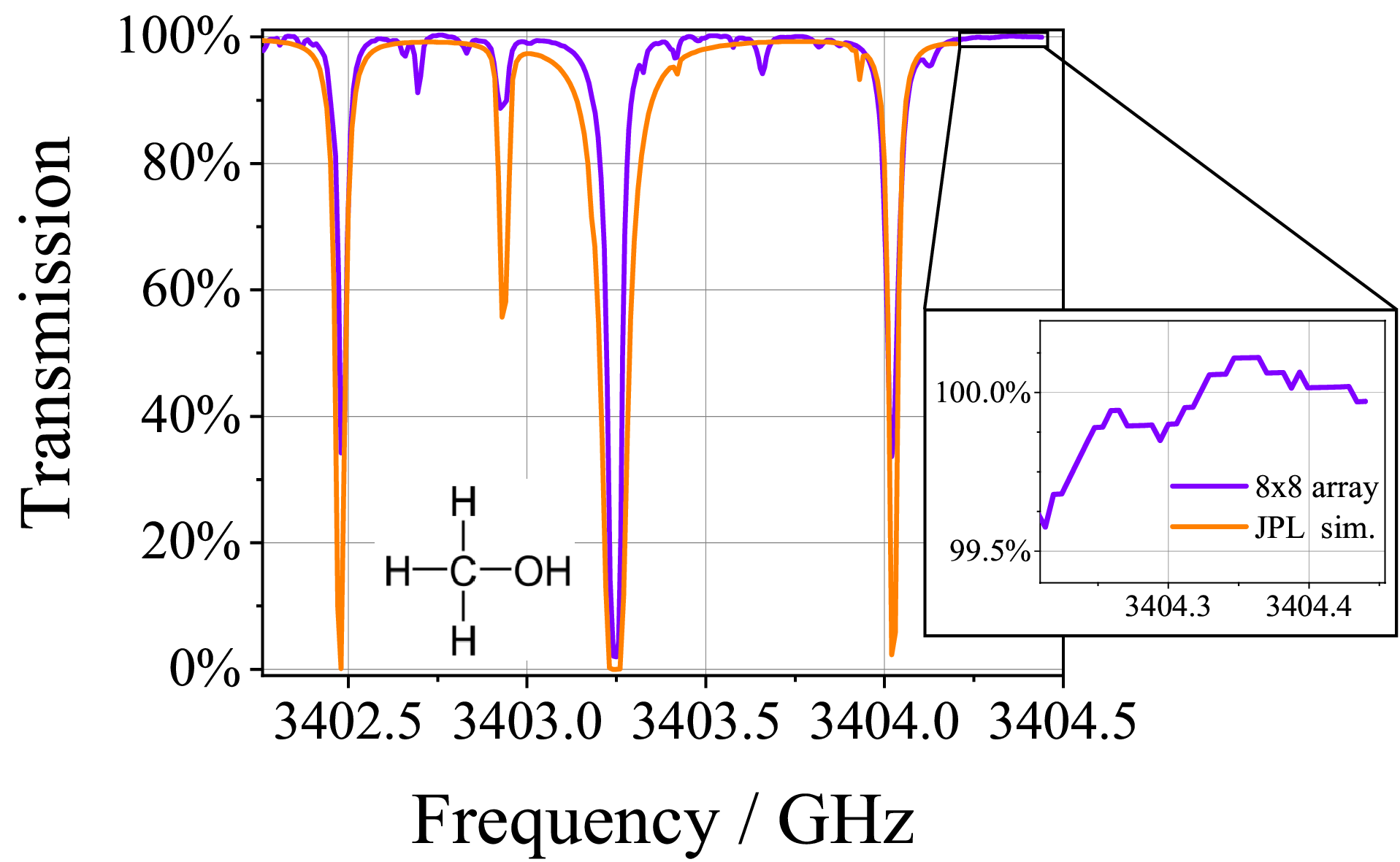}
  \caption{Experimental Methanol transmission spectrum around \SIrange{3.402}{3.405}{\tera \hertz} measured with the 8$\times$8 detector array. The methanol vapor was stored in a \SI{0.95}{\meter} gas cell at a pressure of 0.5~Torr (\SI{66.6}{\pascal}). Corresponding simulated NASA JPL catalogue data available  for methanol vapor is shown for comparison.  
  Inset: expanded view of noise-floor of measurement. }
  \label{fig:MethanolSpectrum}
\end{figure}

Measurements were repeated at QCL-temperatures from 20~K to 55~K, to provide additional tuning of the laser emission, and results were stitched numerically to provide a continuous spectrum.
Finally, the results were normalized to transmission through the empty gas cell, and the observed methanol absorption features were referenced to cataloged data~\cite{pickett_submillimeter_1998} to provide a frequency calibration.
\autoref{fig:MethanolSpectrum} shows the final transmission data, over
a spectroscopic bandwidth of \SIrange{3.402}{3.405}{\tera \hertz} at $\sim\SI{5}{\mega\hertz}$ resolution.
The low detector noise potentially enables a very low molecular detection limit.
Taking the exemplar absorption line at 3402.477~GHz, a transmission of 36\% was observed.
The Beer--Lambert law,
\begin{equation}
    T = \frac{P}{P_0} = \exp(-\sigma n L)
\end{equation}
then provides an equivalent absorption cross-section of $\sigma=\SI{6.5e-19}{\centi \meter \squared\per molecule}$ for this absorption line, where $L=\SI{95}{\centi \meter}$ is the gas-cell length, and $n=\SI{1.6e16}{molecules\per\cm\cubed}$ is the numerical density of gas in the cell.
The latter is determined from the ideal gas law, using
\begin{equation}
    n=\frac{p}{k_B T}
\end{equation}
where $p$ is the gas pressure (in Pa), $k_B$ is the Boltzmann constant and $T=\SI{293}{\kelvin}$ is the gas temperature set by the laboratory climate control.
Pressure $p$ inside the gas-cell was determined by a MKS a-Baratron\texttrademark{}  capacitance manometer specified for the range below 10~Torr (\SI{1333}{\Pa}). As a pressure controller, we used a MKS 946 vacuum system controller.
At the lower detection limit (i.e., with very low absorbance) the Beer--Lambert law is approximated well using a first-order MacLaurin expansion, such that:
\begin{equation}
    \Delta P = -\sigma n L P_0
\end{equation}
where $\Delta P$ is the small drop in THz power transmitted through the cell, resulting from molecular absorption.
The detector noise sets a practical limit on the lowest measurable $\Delta P$.
We approximate this as three times the noise standard deviation, such that
$\Delta P_{min} \approx 3\cdot~\text{NEP}~\cdot~\sqrt{\Delta f}$.

For the integration time used in this work, this results in a minimum detectable absorbance $A_{\text{min}}\approx 3 \text{NEP}\sqrt{\Delta f}/P_0 = \num{1.6e-5}$.
For the selected methanol absorption line, this corresponds to a minimum molecular concentration $n_{min}=\SI{2.6e11}{molecule\per\centi\metre\cubed}$, or a partial pressure of \qty{1.0}{\milli\pascal} methanol vapor within an inert \qty{66.6}{\pascal} balance gas.
\footnote{It is important to note that this absorbance limit assumes that ~1 mW optical power is coupled into the detector (i.e., ~30~\% of the source power).  As such, the practically realisable detection limit will be sensitive to optical alignment. Alternatively, an effective full-system NEP would allow a more accurate practical estimate to be obtained.}

\section{Summary and Conclusion}
In this article, we have presented a fast and sensitive room-temperature detector for the terahertz range \SI{3}{\tera \hertz} and above, which is of major interest for gas spectroscopy applications in combination with powerful terahertz QCLs. 
We demonstrated the significant potential of this combination (TeraFET+QCL) through experimental gas  spectroscopy data shown in \autoref{fig:MethanolSpectrum} investigating methanol vapor. Our current results show good agreement with simulated NASA Jet Propulsion Laboratory (JPL) catalogue data. 
Since JPL data is purely simulation-based, the strength of the absorption lines is associated with uncertainties. Therefore, greater attention should be paid to their spectral positions, where we find good agreement. On closer inspection, additional weak absorption lines were observed in the experimental spectra within two independent measurement runs. 
Their origin will be the focus of our future research activities. Besides further enhancing data quality by fully utilizing the detector modulation bandwidth now available, we aim to examine other important gas species (e.g. ammonium).  
 Employing faster sampling rates could decrease measurement duration or enhance spectral resolution by implementing smaller step widths in the  present  QCL-current biasing technique. 
We intend to leverage our advancements in experimental capabilities to expand our dataset to include other important gases such as atomic oxygen or ammonia. 
All of these experiments benefit from the currently available modulation bandwidth, ranging from the low Hz-range, enabling real-time QCL analysis \cite{north_real-time_2023}, up to $f_{-3dB}\approx\SI{21}{\mega\hertz}$, which can be utilized,  in combination with the high sensitivity for future time-resolved gas tracing experiments. A minimum detectable absorbance of \num{1.6e-5} was estimated.
In case of the methanol vapor lines studied in this work, this corresponds to a minimum molecular concentration of \SI{2.6e11}{molecule\per\centi\metre\cubed}, or a partial pressure of \qty{1.0}{\milli\pascal} within an inert balance gas at \qty{66.6}{\pascal}.
It is important to note that this molecular detection limit relates to the complete system, rather than just the detector.
This is influenced by factors including the THz source power and stability, detector NEP, measurement bandwidth, gas cell length, balance pressure, and the absorption cross-section of the spectroscopic line under study.
In practice, therefore, better molecular sensitivity could be achieved with the detector arrays described in this work, through the use of multi-pass spectroscopy techniques, or improvements in THz source performance.

\section*{Acknowledgments}
The authors gratefully acknowledge A.~Orr--Ewing, University of Bristol, for provision of a gas-spectroscopy cell, and D.~Stone, University of Leeds, for helpful discussions relating to gas spectroscopy. We acknowledge Dmytro B. But from CENTERA Laboratories, Warsaw, Poland for sharing prelimary data on the detector's spectral response.

\section*{Data Availability}
Data associated with this paper are freely available at the Goethe-University Frankfurt Data Repository 
\url{https://doi.org/10.25716/gude.1m9h-e07d}   

\section*{Author contributions}
J.~Holstein: Investigation (lead), Methodology, Visualization, Writing --- original draft;
N.~K.~North: Investigation;
M.~D.~Horbury: Investigation;
S.~S.~Kondawar: Investigation;
I.~Kundu: Investigation;
M.~Salih: Investigation;
A.~Krysl: Investigation;
L.~Li: Investigation, Resources;
E.~H.~Linfield: Conceptualization, Funding Acquisition;
J.~R.~Freeman: Conceptualization, Funding Acquisition;
A.~Valavanis: Conceptualization, Methodology, Investigation, Funding Acquisition, Supervision, Writing --- Review and Editing;
A.~Lisauskas: Conceptualization, Methodology, Funding Acquisition, Supervision;
H.~G.~Roskos: Conceptualization, Methodology, Funding Acquisition, Supervision, Writing.

\ifCLASSOPTIONcaptionsoff
  \newpage
\fi

\bibliography{mainIEEE}

\begin{thebibliography}{10}
\providecommand{\url}[1]{#1}
\csname url@samestyle\endcsname
\providecommand{\newblock}{\relax}
\providecommand{\bibinfo}[2]{#2}
\providecommand{\BIBentrySTDinterwordspacing}{\spaceskip=0pt\relax}
\providecommand{\BIBentryALTinterwordstretchfactor}{4}
\providecommand{\BIBentryALTinterwordspacing}{\spaceskip=\fontdimen2\font plus
\BIBentryALTinterwordstretchfactor\fontdimen3\font minus \fontdimen4\font\relax}
\providecommand{\BIBforeignlanguage}[2]{{%
\expandafter\ifx\csname l@#1\endcsname\relax
\typeout{** WARNING: IEEEtran.bst: No hyphenation pattern has been}%
\typeout{** loaded for the language `#1'. Using the pattern for}%
\typeout{** the default language instead.}%
\else
\language=\csname l@#1\endcsname
\fi
#2}}
\providecommand{\BIBdecl}{\relax}
\BIBdecl

\bibitem{hubers_high-resolution_2019}
\BIBentryALTinterwordspacing
H.-W. Hübers, H.~Richter, and M.~Wienold, ``\BIBforeignlanguage{en}{High-resolution terahertz spectroscopy with quantum-cascade lasers},'' \emph{\BIBforeignlanguage{en}{Journal of Applied Physics}}, vol. 125, no.~15, p. 151401, Apr. 2019. [Online]. Available: \url{http://aip.scitation.org/doi/10.1063/1.5084105}
\BIBentrySTDinterwordspacing

\bibitem{li_terahertz_2014}
\BIBentryALTinterwordspacing
L.~Li, L.~Chen, J.~Zhu, J.~Freeman, P.~Dean, A.~Valavanis, A.~Davies, and E.~Linfield, ``\BIBforeignlanguage{en}{Terahertz quantum cascade lasers with {\textgreater}1 {W} output powers},'' \emph{\BIBforeignlanguage{en}{Electronics Letters}}, vol.~50, no.~4, pp. 309--311, Feb. 2014. [Online]. Available: \url{https://onlinelibrary.wiley.com/doi/10.1049/el.2013.4035}
\BIBentrySTDinterwordspacing

\bibitem{williams_terahertz_2007}
\BIBentryALTinterwordspacing
B.~S. Williams, ``\BIBforeignlanguage{en}{Terahertz quantum-cascade lasers},'' \emph{\BIBforeignlanguage{en}{Nature Photonics}}, vol.~1, no.~9, pp. 517--525, Sep. 2007. [Online]. Available: \url{http://www.nature.com/articles/nphoton.2007.166}
\BIBentrySTDinterwordspacing

\bibitem{richter_direct_2021}
\BIBentryALTinterwordspacing
H.~Richter, C.~Buchbender, R.~Güsten, R.~Higgins, B.~Klein, J.~Stutzki, H.~Wiesemeyer, and H.-W. Hübers, ``\BIBforeignlanguage{en}{Direct measurements of atomic oxygen in the mesosphere and lower thermosphere using terahertz heterodyne spectroscopy},'' \emph{\BIBforeignlanguage{en}{Communications Earth \& Environment}}, vol.~2, no.~1, p.~19, Jan. 2021. [Online]. Available: \url{https://www.nature.com/articles/s43247-020-00084-5}
\BIBentrySTDinterwordspacing

\bibitem{wubs_terahertz_2023}
\BIBentryALTinterwordspacing
J.~R. Wubs, U.~Macherius, K.-D. Weltmann, X.~Lü, B.~Röben, K.~Biermann, L.~Schrottke, H.~T. Grahn, and J.~H. Van~Helden, ``Terahertz absorption spectroscopy for measuring atomic oxygen densities in plasmas,'' \emph{Plasma Sources Science and Technology}, vol.~32, no.~2, p. 025006, Feb. 2023. [Online]. Available: \url{https://iopscience.iop.org/article/10.1088/1361-6595/acb815}
\BIBentrySTDinterwordspacing

\bibitem{cuisset_terahertz_2021}
\BIBentryALTinterwordspacing
A.~Cuisset, F.~Hindle, G.~Mouret, R.~Bocquet, J.~Bruckhuisen, J.~Decker, A.~Pienkina, C.~Bray, E.~Fertein, and V.~Boudon, ``\BIBforeignlanguage{en}{Terahertz {Rotational} {Spectroscopy} of {Greenhouse} {Gases} {Using} {Long} {Interaction} {Path}-{Lengths}},'' \emph{\BIBforeignlanguage{en}{Applied Sciences}}, vol.~11, no.~3, p. 1229, Jan. 2021. [Online]. Available: \url{https://www.mdpi.com/2076-3417/11/3/1229}
\BIBentrySTDinterwordspacing

\bibitem{dean_terahertz_2014}
\BIBentryALTinterwordspacing
P.~Dean, A.~Valavanis, J.~Keeley, K.~Bertling, Y.~L. Lim, R.~Alhathlool, A.~D. Burnett, L.~H. Li, S.~P. Khanna, D.~Indjin, T.~Taimre, A.~D. Rakić, E.~H. Linfield, and A.~G. Davies, ``Terahertz imaging using quantum cascade lasers—a review of systems and applications,'' \emph{Journal of Physics D: Applied Physics}, vol.~47, no.~37, p. 374008, Sep. 2014. [Online]. Available: \url{https://iopscience.iop.org/article/10.1088/0022-3727/47/37/374008}
\BIBentrySTDinterwordspacing

\bibitem{vitiello_terahertz_2022}
\BIBentryALTinterwordspacing
M.~S. Vitiello and P.~De~Natale, ``\BIBforeignlanguage{en}{Terahertz {Quantum} {Cascade} {Lasers} as {Enabling} {Quantum} {Technology}},'' \emph{\BIBforeignlanguage{en}{Advanced Quantum Technologies}}, vol.~5, no.~1, p. 2100082, Jan. 2022. [Online]. Available: \url{https://onlinelibrary.wiley.com/doi/10.1002/qute.202100082}
\BIBentrySTDinterwordspacing

\bibitem{gao_recent_2023}
\BIBentryALTinterwordspacing
L.~Gao, C.~Feng, and X.~Zhao, ``\BIBforeignlanguage{en}{Recent developments in terahertz quantum cascade lasers for practical applications},'' \emph{\BIBforeignlanguage{en}{Nanotechnology Reviews}}, vol.~12, no.~1, p. 20230115, Sep. 2023. [Online]. Available: \url{https://www.degruyter.com/document/doi/10.1515/ntrev-2023-0115/html}
\BIBentrySTDinterwordspacing

\bibitem{javadi_sensitivity_2021}
\BIBentryALTinterwordspacing
E.~Javadi, D.~B. But, K.~Ikamas, J.~Zdanevičius, W.~Knap, and A.~Lisauskas, ``\BIBforeignlanguage{en}{Sensitivity of {Field}-{Effect} {Transistor}-{Based} {Terahertz} {Detectors}},'' \emph{\BIBforeignlanguage{en}{Sensors}}, vol.~21, no.~9, p. 2909, Apr. 2021. [Online]. Available: \url{https://www.mdpi.com/1424-8220/21/9/2909}
\BIBentrySTDinterwordspacing

\bibitem{regensburger_picosecond-scale_2019}
\BIBentryALTinterwordspacing
S.~Regensburger, S.~Winnerl, J.~M. Klopf, H.~Lu, A.~C. Gossard, and S.~Preu, ``Picosecond-{Scale} {Terahertz} {Pulse} {Characterization} {With} {Field}-{Effect} {Transistors},'' \emph{IEEE Transactions on Terahertz Science and Technology}, vol.~9, no.~3, pp. 262--271, May 2019. [Online]. Available: \url{https://ieeexplore.ieee.org/document/8662700/}
\BIBentrySTDinterwordspacing

\bibitem{yadav_state---art_2023}
\BIBentryALTinterwordspacing
R.~Yadav, F.~Ludwig, F.~R. Faridi, J.~M. Klopf, H.~G. Roskos, S.~Preu, and A.~Penirschke, ``\BIBforeignlanguage{en}{State-of-the-{Art} {Room} {Temperature} {Operable} {Zero}-{Bias} {Schottky} {Diode}-{Based} {Terahertz} {Detector} {Up} to 5.56 {THz}},'' \emph{\BIBforeignlanguage{en}{Sensors}}, vol.~23, no.~7, p. 3469, Mar. 2023. [Online]. Available: \url{https://www.mdpi.com/1424-8220/23/7/3469}
\BIBentrySTDinterwordspacing

\bibitem{zdanevicius_field-effect_2018}
\BIBentryALTinterwordspacing
J.~Zdanevicius, D.~Cibiraite, K.~Ikamas, M.~Bauer, J.~Matukas, A.~Lisauskas, H.~Richter, T.~Hagelschuer, V.~Krozer, H.-W. Hubers, and H.~G. Roskos, ``Field-{Effect} {Transistor} {Based} {Detectors} for {Power} {Monitoring} of {THz} {Quantum} {Cascade} {Lasers},'' \emph{IEEE Transactions on Terahertz Science and Technology}, vol.~8, no.~6, pp. 613--621, Nov. 2018. [Online]. Available: \url{https://ieeexplore.ieee.org/document/8536434/}
\BIBentrySTDinterwordspacing

\bibitem{ikamas_efficient_2017}
\BIBentryALTinterwordspacing
K.~Ikamas, A.~Lisauskas, S.~Boppel, Q.~Hu, and H.~G. Roskos, ``\BIBforeignlanguage{en}{Efficient {Detection} of 3 {THz} {Radiation} from {Quantum} {Cascade} {Laser} {Using} {Silicon} {CMOS} {Detectors}},'' \emph{\BIBforeignlanguage{en}{Journal of Infrared, Millimeter, and Terahertz Waves}}, vol.~38, no.~10, pp. 1183--1188, Oct. 2017. [Online]. Available: \url{http://link.springer.com/10.1007/s10762-017-0407-9}
\BIBentrySTDinterwordspacing

\bibitem{wiecha_antenna-coupled_2021}
\BIBentryALTinterwordspacing
M.~M. Wiecha, R.~Kapoor, A.~V. Chernyadiev, K.~Ikamas, A.~Lisauskas, and H.~G. Roskos, ``\BIBforeignlanguage{en}{Antenna-coupled field-effect transistors as detectors for terahertz near-field microscopy},'' \emph{\BIBforeignlanguage{en}{Nanoscale Advances}}, vol.~3, no.~6, pp. 1717--1724, 2021. [Online]. Available: \url{http://xlink.rsc.org/?DOI=D0NA00928H}
\BIBentrySTDinterwordspacing

\bibitem{cibiraite-lukenskiene_passive_2020}
\BIBentryALTinterwordspacing
D.~Čibiraitė Lukenskienė, K.~Ikamas, T.~Lisauskas, V.~Krozer, H.~G. Roskos, and A.~Lisauskas, ``\BIBforeignlanguage{en}{Passive {Detection} and {Imaging} of {Human} {Body} {Radiation} {Using} an {Uncooled} {Field}-{Effect} {Transistor}-{Based} {THz} {Detector}},'' \emph{\BIBforeignlanguage{en}{Sensors}}, vol.~20, no.~15, p. 4087, Jul. 2020. [Online]. Available: \url{https://www.mdpi.com/1424-8220/20/15/4087}
\BIBentrySTDinterwordspacing

\bibitem{zak_antenna-integrated_2014}
\BIBentryALTinterwordspacing
A.~Zak, M.~A. Andersson, M.~Bauer, J.~Matukas, A.~Lisauskas, H.~G. Roskos, and J.~Stake, ``\BIBforeignlanguage{en}{Antenna-{Integrated} 0.6 {THz} {FET} {Direct} {Detectors} {Based} on {CVD} {Graphene}},'' \emph{\BIBforeignlanguage{en}{Nano Letters}}, vol.~14, no.~10, pp. 5834--5838, Oct. 2014. [Online]. Available: \url{https://pubs.acs.org/doi/10.1021/nl5027309}
\BIBentrySTDinterwordspacing

\bibitem{generalov_400-ghz_2017}
\BIBentryALTinterwordspacing
A.~A. Generalov, M.~A. Andersson, X.~Yang, A.~Vorobiev, and J.~Stake, ``\BIBforeignlanguage{en}{A 400-{GHz} {Graphene} {FET} {Detector}},'' \emph{\BIBforeignlanguage{en}{IEEE Transactions on Terahertz Science and Technology}}, vol.~7, no.~5, pp. 614--616, Sep. 2017. [Online]. Available: \url{http://ieeexplore.ieee.org/document/7981343/}
\BIBentrySTDinterwordspacing

\bibitem{ludwig_terahertz_2024}
\BIBentryALTinterwordspacing
F.~Ludwig, A.~Generalov, J.~Holstein, A.~Murros, K.~Viisanen, M.~Prunnila, and H.~G. Roskos, ``\BIBforeignlanguage{en}{Terahertz {Detection} with {Graphene} {FETs}: {Photothermoelectric} and {Resistive} {Self}-{Mixing} {Contributions} to the {Detector} {Response}},'' \emph{\BIBforeignlanguage{en}{ACS Applied Electronic Materials}}, p. acsaelm.3c01511, Mar. 2024. [Online]. Available: \url{https://pubs.acs.org/doi/10.1021/acsaelm.3c01511}
\BIBentrySTDinterwordspacing

\bibitem{bauer_high-sensitivity_2019}
\BIBentryALTinterwordspacing
M.~Bauer, A.~Ramer, S.~A. Chevtchenko, K.~Y. Osipov, D.~Cibiraite, S.~Pralgauskaite, K.~Ikamas, A.~Lisauskas, W.~Heinrich, V.~Krozer, and H.~G. Roskos, ``\BIBforeignlanguage{en}{A {High}-{Sensitivity} {AlGaN}/{GaN} {HEMT} {Terahertz} {Detector} {With} {Integrated} {Broadband} {Bow}-{Tie} {Antenna}},'' \emph{\BIBforeignlanguage{en}{IEEE Transactions on Terahertz Science and Technology}}, vol.~9, no.~4, pp. 430--444, Jul. 2019. [Online]. Available: \url{https://ieeexplore.ieee.org/document/8734725/}
\BIBentrySTDinterwordspacing

\bibitem{boppel_cmos_2012}
\BIBentryALTinterwordspacing
S.~Boppel, A.~Lisauskas, M.~Mundt, D.~Seliuta, L.~Minkevicius, I.~Kasalynas, G.~Valusis, M.~Mittendorff, S.~Winnerl, V.~Krozer, and H.~G. Roskos, ``\BIBforeignlanguage{en}{{CMOS} {Integrated} {Antenna}-{Coupled} {Field}-{Effect} {Transistors} for the {Detection} of {Radiation} {From} 0.2 to 4.3 {THz}},'' \emph{\BIBforeignlanguage{en}{IEEE Transactions on Microwave Theory and Techniques}}, vol.~60, no.~12, pp. 3834--3843, Dec. 2012. [Online]. Available: \url{http://ieeexplore.ieee.org/document/6353608/}
\BIBentrySTDinterwordspacing

\bibitem{bauer_antenna-coupled_2014}
\BIBentryALTinterwordspacing
M.~Bauer, R.~Venckevičius, I.~Kašalynas, S.~Boppel, M.~Mundt, L.~Minkevičius, A.~Lisauskas, G.~Valušis, V.~Krozer, and H.~G. Roskos, ``\BIBforeignlanguage{en}{Antenna-coupled field-effect transistors for multi-spectral terahertz imaging up to 425 {THz}},'' \emph{\BIBforeignlanguage{en}{Optics Express}}, vol.~22, no.~16, p. 19235, Aug. 2014. [Online]. Available: \url{https://opg.optica.org/oe/abstract.cfm?uri=oe-22-16-19235}
\BIBentrySTDinterwordspacing

\bibitem{ludwig_modeling_2024}
\BIBentryALTinterwordspacing
F.~Ludwig, J.~Holstein, A.~Krysl, A.~Lisauskas, and H.~G. Roskos, ``Modeling of {Antenna}-{Coupled} {Si} {MOSFETs} in the {Terahertz} {Frequency} {Range},'' \emph{IEEE Transactions on Terahertz Science and Technology}, vol.~14, no.~3, pp. 414--423, May 2024. [Online]. Available: \url{https://ieeexplore.ieee.org/document/10497872/}
\BIBentrySTDinterwordspacing

\bibitem{ikamas_broadband_2018}
\BIBentryALTinterwordspacing
K.~Ikamas, D.~Cibiraite, A.~Lisauskas, M.~Bauer, V.~Krozer, and H.~G. Roskos, ``Broadband {Terahertz} {Power} {Detectors} {Based} on 90-nm {Silicon} {CMOS} {Transistors} {With} {Flat} {Responsivity} {Up} to 2.2 {THz},'' \emph{IEEE Electron Device Letters}, vol.~39, no.~9, pp. 1413--1416, Sep. 2018. [Online]. Available: \url{https://ieeexplore.ieee.org/document/8418740/}
\BIBentrySTDinterwordspacing

\bibitem{ikamas_all-electronic_2021}
\BIBentryALTinterwordspacing
K.~Ikamas, D.~B. But, A.~Cesiul, C.~Kołaciński, T.~Lisauskas, W.~Knap, and A.~Lisauskas, ``\BIBforeignlanguage{en}{All-{Electronic} {Emitter}-{Detector} {Pairs} for 250 {GHz} in {Silicon}},'' \emph{\BIBforeignlanguage{en}{Sensors}}, vol.~21, no.~17, p. 5795, Aug. 2021. [Online]. Available: \url{https://www.mdpi.com/1424-8220/21/17/5795}
\BIBentrySTDinterwordspacing

\bibitem{dyakonov_shallow_1993}
\BIBentryALTinterwordspacing
M.~Dyakonov and M.~Shur, ``\BIBforeignlanguage{en}{Shallow water analogy for a ballistic field effect transistor: {New} mechanism of plasma wave generation by dc current},'' \emph{\BIBforeignlanguage{en}{Physical Review Letters}}, vol.~71, no.~15, pp. 2465--2468, Oct. 1993. [Online]. Available: \url{https://link.aps.org/doi/10.1103/PhysRevLett.71.2465}
\BIBentrySTDinterwordspacing

\bibitem{dyakonov_detection_1996}
------, ``Detection, mixing, and frequency multiplication of terahertz radiation by two-dimensional electronic fluid,'' \emph{IEEE Transactions on Electron Devices}, vol.~43, no.~3, pp. 380--387, 1996.

\bibitem{horbury_real-time_2023}
\BIBentryALTinterwordspacing
M.~D. Horbury, N.~K. North, J.~Holstein, H.~Godden, L.~H. Li, J.~R. Freeman, E.~H. Linfield, H.~Roskos, A.~Lisauskas, and A.~Valavanis, ``Real-{Time} {Terahertz} {Absorption} {Spectroscopy} of {Methanol} and {Deuterated}-{Methanol} {Vapour}, using a {TeraFET} {Detector} {Array},'' in \emph{2023 48th {International} {Conference} on {Infrared}, {Millimeter}, and {Terahertz} {Waves} ({IRMMW}-{THz})}.\hskip 1em plus 0.5em minus 0.4em\relax Montreal, QC, Canada: IEEE, Sep. 2023, pp. 1--2. [Online]. Available: \url{https://ieeexplore.ieee.org/document/10298916/}
\BIBentrySTDinterwordspacing

\bibitem{north_real-time_2023}
\BIBentryALTinterwordspacing
N.~K. North, J.~Holstein, M.~D. Horbury, H.~Godden, L.~H. Li, J.~R. Freeman, E.~H. Linfield, H.~Roskos, A.~Lisauskas, and A.~Valavanis, ``Real-{Time} {Analysis} of {THz} {Quantum}-{Cascade} {Laser} {Signals} using a {Field} {Effect} {Transistor} {Array},'' in \emph{2023 48th {International} {Conference} on {Infrared}, {Millimeter}, and {Terahertz} {Waves} ({IRMMW}-{THz})}.\hskip 1em plus 0.5em minus 0.4em\relax Montreal, QC, Canada: IEEE, Sep. 2023, pp. 1--2. [Online]. Available: \url{https://ieeexplore.ieee.org/document/10298938/}
\BIBentrySTDinterwordspacing

\bibitem{krysl_control_2022}
\BIBentryALTinterwordspacing
A.~Krysl, D.~B. But, K.~Ikamas, H.~Yuan, M.~Kocybik, M.~Bauer, F.~Friederich, A.~Lisauskas, and H.~G. Roskos, ``Control and {Optimization} of {Patch}-{Antenna}-{Coupled} {THz} {Detector} {Performance} using {Superstrate} {Dielectric} and {Silicon} {Lens},'' in \emph{2022 47th {International} {Conference} on {Infrared}, {Millimeter} and {Terahertz} {Waves} ({IRMMW}-{THz})}.\hskip 1em plus 0.5em minus 0.4em\relax Delft, Netherlands: IEEE, Aug. 2022, pp. 1--2. [Online]. Available: \url{https://ieeexplore.ieee.org/document/9896100/}
\BIBentrySTDinterwordspacing

\bibitem{linear_systems_lsk389_nodate}
\BIBentryALTinterwordspacing
{Linear Systems}, ``{LSK389} {A}/{B}/{C}/{D} {Ultra}-{Low} {Noise} {Monolithic} {Dual} {N}-{Channel} {JFET} {Amplifier}.'' [Online]. Available: \url{https://www.linearsystems.com/_files/ugd/7e8069_c90834b817214602abcbf263293295f2.pdf}
\BIBentrySTDinterwordspacing

\bibitem{brown_fundamentals_2003}
\BIBentryALTinterwordspacing
E.~R. Brown, ``\BIBforeignlanguage{en}{Fundamentals of {Terrestrial} {Millimeter}-{Wave} and {THz} {Remote} {Sensing}},'' \emph{\BIBforeignlanguage{en}{International Journal of High Speed Electronics and Systems}}, vol.~13, no.~04, pp. 995--1097, Dec. 2003. [Online]. Available: \url{https://www.worldscientific.com/doi/abs/10.1142/S0129156403002125}
\BIBentrySTDinterwordspacing

\bibitem{wienold_high-temperature_2014}
\BIBentryALTinterwordspacing
M.~Wienold, B.~Röben, L.~Schrottke, R.~Sharma, A.~Tahraoui, K.~Biermann, and H.~T. Grahn, ``\BIBforeignlanguage{EN}{High-temperature, continuous-wave operation of terahertz quantum-cascade lasers with metal-metal waveguides and third-order distributed feedback},'' \emph{\BIBforeignlanguage{EN}{Optics Express}}, vol.~22, no.~3, pp. 3334--3348, Feb. 2014, publisher: Optica Publishing Group. [Online]. Available: \url{https://opg.optica.org/oe/abstract.cfm?uri=oe-22-3-3334}
\BIBentrySTDinterwordspacing

\bibitem{wienold_low-voltage_2009}
M.~Wienold, L.~Schrottke, M.~Giehler, R.~Hey, W.~Anders, and H.~Grahn, ``Low-voltage terahertz quantum-cascade lasers based on {LO}-phonon-assisted interminiband transitions,'' \emph{Electronics Letters}, vol.~45, no.~20, pp. 1030--1031, Sep. 2009.

\bibitem{gellie_injection-locking_2010}
\BIBentryALTinterwordspacing
P.~Gellie, S.~Barbieri, J.-F. Lampin, P.~Filloux, C.~Manquest, C.~Sirtori, I.~Sagnes, S.~P. Khanna, E.~H. Linfield, A.~G. Davies, H.~Beere, and D.~Ritchie, ``\BIBforeignlanguage{en}{Injection-locking of terahertz quantum cascade lasers up to {35GHz} using {RF} amplitude modulation},'' \emph{\BIBforeignlanguage{en}{Optics Express}}, vol.~18, no.~20, p. 20799, Sep. 2010. [Online]. Available: \url{https://opg.optica.org/abstract.cfm?URI=oe-18-20-20799}
\BIBentrySTDinterwordspacing

\bibitem{pickett_submillimeter_1998}
\BIBentryALTinterwordspacing
H.~M. Pickett, R.~L. Poynter, E.~A. Cohen, M.~L. Delitsky, J.~C. Pearson, and H.~S.~P. Müller, ``Submillimeter, millimeter, and microwave spectral line catalog,'' \emph{Journal of Quantitative Spectroscopy and Radiative Transfer}, vol.~60, no.~5, pp. 883--890, Nov. 1998. [Online]. Available: \url{http://www.sciencedirect.com/science/article/pii/S0022407398000910}
\BIBentrySTDinterwordspacing

\end{thebibliography}
\bibliographystyle{IEEEtran}

\end{document}